\documentclass
[aps,prl,twocolumn,floatfix,english,showpacs,10pt,superscriptaddress]{revtex4-2}%
\usepackage{graphicx}
\usepackage{booktabs}
\usepackage{amsmath}
\usepackage{physics}
\usepackage{amssymb}
\usepackage{colordvi}
\usepackage{verbatim}
\usepackage{xcolor}
\usepackage{mathrsfs}
\usepackage{epsfig}
\usepackage{lipsum}
\usepackage{amsfonts}
\usepackage{makecell}
\usepackage{esint}
\usepackage{bm}
\usepackage[most]{tcolorbox}
\usepackage[normalem]{ulem}
\usepackage[unicode=true, breaklinks=false, pdfborder={0 0 1}, backref=false,
colorlinks=true, linkcolor=blue, urlcolor=blue, citecolor=blue]{hyperref}%
\providecommand{\U}[1]{\protect\rule{.1in}{.1in}}
\providecommand{\U}[1]{\protect\rule{.1in}{.1in}}
\setcitestyle{numbers,square}

\begin{document}

\title{Structural Chirality Switches Quantum Coherence of Magnons}

\author{Yanmeng Lei}
\affiliation{School of Physics, Huazhong University of Science and Technology, Wuhan 430074, China}

\author{Ji Zou}
\affiliation{Physics Department, King Fahd University of Petroleum and Minerals, 31261, Dhahran, Saudi Arabia}
\affiliation{Quantum Center, KFUPM, Dhahran, Saudi Arabia}

\author{Sanchar Sharma}
\affiliation{Departamento de F\'isica Te\'orica de la Materia Condensada, Universidad Aut\'onoma de Madrid, E-28049 Madrid, Spain}
\affiliation{Condensed Matter Physics Center (IFIMAC), Universidad Aut\'onoma de Madrid, E-28049 Madrid, Spain}

\author{Tao Yu}
\email{taoyuphy@hust.edu.cn}
\affiliation{School of Physics, Huazhong University of Science and Technology, Wuhan 430074, China}

\date{\today }

\begin{abstract}
Structural chirality is ubiquitous in nature, yet its impact on spin transport and quantum resources remains largely elusive. Here we establish the relation between chiral spin transport and quantum coherence by predicting the structural-chirality-switched response of both ferromagnetic and antiferromagnetic spirals in the electric spin injection-detection configuration. Such spirals are excellent for quantum coherence since their screw symmetry forbids any nonzero diagonal element of the magnon spin matrix: every eigenmode is linearly polarized and carries no spin, so spin transport is mediated entirely by quantum coherence between distinct modes.  Reversing chirality flips the magnon spin along the spiral axis and the injected quantum coherence in lockstep, leaving the associated injected spin current invariant. Spiral chirality acts instead at the detection interface, where electron--magnon scattering amplitude is reversed and causes the magnon accumulation to be read out as a voltage of opposite sign.  This interfacial readout is consistent with reversed first-harmonic voltage under electric-field switching in multiferroic BiFeO$_3$, establishing structural chirality as an intrinsic control parameter for magnon quantum coherence and its electrical readout.

\end{abstract}

\maketitle

\textit{Introduction}.---Chirality, the property of being non-superimposable on one's mirror image, is broadly classified into dynamical and structural forms, which play a fundamental role in quantum transport and quantum resources with materials. The locking of circular polarization and propagation direction of waves~\cite{chiral_phonon1,chiral_phonon2,chiral_phonon3,chiral_phonon4,chiral_magnon1,chiral_magnon2,chiral_magnon3}, or dynamical chirality, is universal~\cite {qing2026,puebla2026}, leading to chiral wave-matter interactions~\cite{wave1,wave2,wave3,wave4,wave5,wave6,wave7,wave8} and nonreciprocal transport~\cite{transport1,transport2,transport3,transport4}. 
Structural chirality introduces a static, geometric paradigm. A striking example is chirality-induced spin selectivity (CISS):  electrons transmitted through chiral molecules acquire spin polarization depending on the molecular chirality~\cite{CISS1,CISS2,CISS3,CISS4,CISS5,CISS6}. Magnetic systems host ubiquitous chiral spin textures such as magnetic skyrmions~\cite{sky1,sky2,sky3,sky4,sky5}, domain walls, and spirals~\cite{spiral1,spiral2,spiral3,spiral4,spiral5,spiral6}, but their effect on spin transport as well as quantum resources is largely unknown (exceptions are magnon scattering by skyrmions~\cite{denisov_skyrmion_2016,iwasaki_magnon_skyrmion_2014,schuette_magnon_skyrmion_2014}).

Structural chirality in magnets describes how magnetic moments rotate in real space and is often governed by the Dzyaloshinskii-Moriya interaction (DMI), with $D>0$ or $D<0$~\cite{DMI1,DMI2}. In typical multiferroic BiFeO$_3$ and its derivatives, the sign of DMI and thereby the structural chirality of underlying magnetic spirals can be switched by an electric field, bringing opportunities of electric control of magnon transport. 
Indeed, recent experiments~\cite{yamasaki_PRL_2007,Ramesh_PRL_2022,Ramesh_NM_2024,Ramesh_NC_2024,Ramesh_AM_2024,Ramesh_npj_2025,Ramesh_PRL_2025,NC-BFO-torque} observed reversed non-local voltages~\cite{nonlocal} mediated by magnon transport via switching $D$, which is interpreted as reversed spin polarization in bulk by the switched weak magnetization~\cite{Ramesh_PRL_2022,Ramesh_AM_2024,Ramesh_PRL_2025}. The electrically driven magnon transport, measured as the first harmonic of voltage~\cite{Ramesh_AM_2024}, differs from thermal transport. A clear microscopic understanding of the relation between structural chirality and magnon quantum transport remains wanting. Further, whether the structural chirality has a useful relation with the quantum property of magnons is unknown.

\begin{figure}[htbp]
    \centering
    \includegraphics[width=8.45cm]{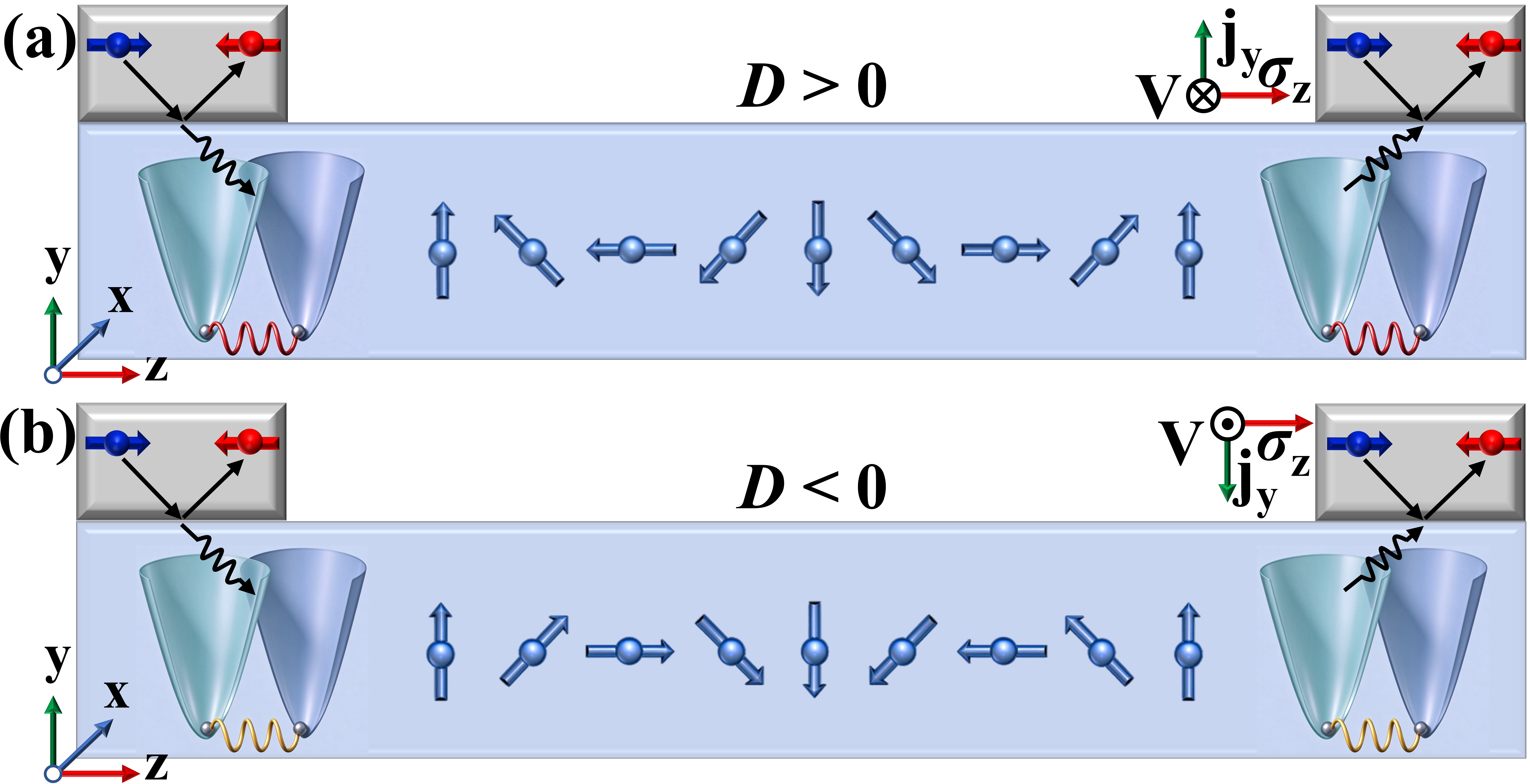}
    \caption{Switched non-local spin-transport of spiral magnets by $D>0$ [(a)] and $D<0$ [(b)]. A charge current in the left normal-metal contact generates electron spin accumulation $\Delta{\pmb \mu}\parallel\hat{\bf z}$ via the spin Hall effect. Through interfacial electron--magnon scattering, the spin flip of electrons from spin-up to spin-down induces opposite quantum coherence between distinct modes in (a) and (b). These quantum coherences diffuse rightward, accumulate, and induce an opposite electron spin current into the right contact via the inverse process.}
    \label{fig1}
\end{figure}

In this Letter, we find that the chiral spin transport is locked with and enabled by the quantum coherence between different magnon modes in both ferromagnetic and antiferromagnetic spirals. We reveal that magnetic spirals are useful for exploiting quantum coherence since the screw symmetry of the spiral texture imposes a selection rule on the magnon spins: all individual magnon modes are linearly polarized and thus carry no spin by modes; instead, magnon spin is carried by a selected quantum coherence across different modes. While reversing the spiral chirality flips the sign of magnon spin along the spiral axis, it simultaneously reverses the injected magnon quantum coherence, leaving the net injected spin current (polarized along the spiral) invariant. Emphasizing the chiral response as a surface effect, we further demonstrate that interfacial electron-magnon scattering is chirality-dependent, such that during detection it converts the magnon accumulation into an inverted voltage signal [see Fig.~\ref{fig1}(a) and (b)], viz., the reversed non-local voltage does not stem from bulk transport modulation.

\textit{Minimal Model}.---The G-type antiferromagnet (AFM) BiFeO$_3$ provides a basis for a minimal model. BiFeO$_3$ consists of two antiferromagnetically coupled spiral sublattices, which in summation maps onto an effective ferromagnetic (FM) spiral~\cite{Gross2017}. To obtain a complete understanding, we consider the AFM spiral~\cite{Gitgeatpong2017} on an equal footing. We isolate the role of magnetic structural chirality by considering a minimal two-dimensional spiral magnet with lattice constant $\delta$,  described by Heisenberg exchange $J$ and DMI~\cite{spiral4,Chen2015Dissipationless}:
\begin{align}	   
\hat{H}=J\sum_{l,l'}\sum_{\langle i,j\rangle=1}^{N}\hat{\bf S}_{l,i}\cdot \hat{\bf S}_{l',j}+\sum_{l,l'}\sum_{\langle i,j\rangle=1}^{N}{\bf D}\cdot\left(\hat{\bf S}_{l,i}\times\hat{\bf S}_{l',j}\right).
\label{Hamiltonian}
\end{align}
Here, $\hat{\bf S}_{l,i}$ represents the atomic spin at the $i$-th sublattice in the $l$-th unit cell, the DMI vector ${\bf D}=\pm D\hat{\bf x}$ for $j_z-i_z=\pm1~(0 ~\rm{otherwise})$, and the summation over the nearest-neighboring sites $\langle\cdots\rangle$ is implied.
The competition between $J$ and $D$ stabilizes a spiral order along $\hat{\bf z}$, while $J<0$ (FM) and $J>0$ (AFM) lead to different alignments of neighboring spins along $\hat{\bf x}$ (Fig.~\ref{fig1} for $J<0$).
The spiral pitch angle $\phi$ is parameterized by $\cos{\phi}=-J/{K}$ and $\sin{\phi}={D}/{K}$ with an  energy scale $K=\sqrt{{J}^{2}+D^{2}}$.
For the FM spiral, $\cos\phi>0$, while $\cos\phi<0$ for AFM. A commensurate spiral has a period of integer $N$ sites, so the pitch angle between neighboring spins is $\phi=2m\pi/N$, with $m=1$ for FM coupling, while $N/4<m<N/2$ and ${\rm gcd}(m,N)=1$ for AFM coupling [see Supplementary Material (SM) for details~\cite{supplement}].
The spiral/cycloid chirality is determined by $\sin\phi=D/K$ and characterized by a chirality index $\kappa={\rm sgn }\left(\hat{\bf x}\cdot\left({\bf S}_{l,i}\times{\bf S}_{l,i+1}\right)\right)$. Reversing the DMI flips $\phi$ to $-\phi$ as well as the chirality $\kappa$ [Fig.~\ref{fig1}(a) and (b)].

\begin{figure}[htp!]
\centering
\includegraphics[width=8.6cm]{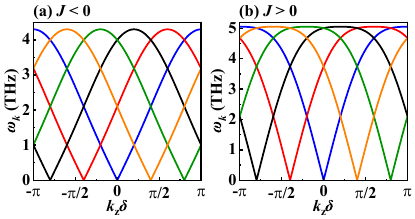}
\caption{Illustrated magnon dispersion for $N=5$, calculated using parameters $K=\sqrt{J^2+D^2}=0.35$~\text{meV}, $S=5/2$, and $\delta=1$~nm. (a): $J<0$ with $\phi=2\pi/N$. (b): $J>0$ with $\phi=4\pi/N$. $|D|=K\sin\phi$ and $J=-K\cos\phi$.}
\label{dispersion}
\end{figure}

Performing the Holstein-Primakoff (HP) transformation renders the Hamiltonian~\eqref{Hamiltonian} expressed as $\hat{H}=(1/2)\sum_{\bf k}\hat{X}_{\bf k}^{\dagger}{\cal H}_{\bf k}\hat{X}_{\bf k}$, where $\hat{X}_{\bf k}=(\hat{a}_{i,\bf k},\hat{a}_{i,-\bf k}^{\dagger})^T$ is composed by HP bosons $\hat{a}_{i,{\bf k}}$ ($\textit{i} = \{1,\cdots,N\}$) and ${\cal H}_{\bf k}$ is the associated $2N\times2N$ matrix.
A Bogoliubov transformation $\hat{X}_{\bf k} = {\cal T}_{\bf k}\hat{\Psi}_{\bf k}$, where $\hat{\Psi}_{\bf k} = (\hat{b}_{i,\bf k},\hat{b}_{i,-\bf k}^\dagger)^T$ collects magnon operators $\hat{b}_{i,\bf k}$, diagonalizes ${\cal H}_{\bf k}$ and yields $\xi=\{1,2,\cdots,N\}$ magnon band with dispersions $\hbar\omega_{\xi,\bf k}=\sqrt{A_\xi^2({\bf k})-B_\xi^2({\bf k})}$, 
with $A_\xi({\bf k})=-KS(\cos\phi+1)\cos(k_z\delta+2\pi (\xi-1)/N)-2KS(\cos\phi\cos(k_x\delta)-\cos\phi-1)$ and $B_\xi({\bf k})=-KS(\cos\phi-1)\cos(k_z\delta+2\pi (\xi-1)/N)$ for FM spiral (see SM~\cite{supplement} for AFM spiral).
As illustrated in Fig.~\ref{dispersion} with $N=5$, the FM and AFM spirals exhibit analogous band structures, as compared in Fig.~\ref{dispersion}(a) and (b). The magnon spectrum exhibits multiple branches originating from band folding, associated with $N$ sublattices in one magnetic unit cell. 
Crucially, these dispersion relations depend only on $\cos\phi$ and thus remain invariant under the reversal of structural chirality ($D \to -D$). This observation rules out the non-local voltage reversal arising from a modified magnon dispersion. Moreover, the multi-branch frequency differences ($\omega_{\xi,\mathbf{q}}-\omega_{\xi',\mathbf{q}}$) govern the quantum coherence, driving a spatial oscillation beat of diffusive spin current, as discussed below.

\begin{figure*}[htbp]
    \centering
    \includegraphics[width=13.5cm]{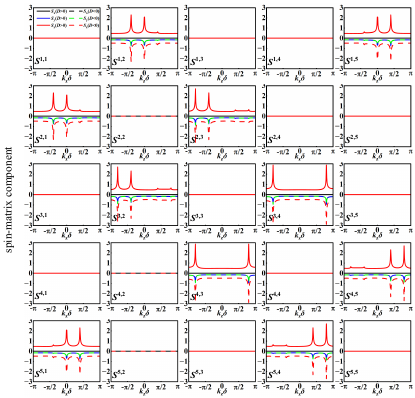}
    \caption{Illustrated magnon spin matrix of FM spiral for $N=5$ with opposite structural chirality, calculated with parameters $K=0.35~{\rm meV}$, $J=-K\cos\phi$, $|D|=K\sin\phi$, and $\delta=1~{\rm nm}$.}
    \label{fig3}
\end{figure*}

\textit{Spin structure of magnons}.---The quadratic terms of magnon operators contribute to the spin operator according to $\hat{\bf S}\rightarrow (1/2)\sum_{\bf k}\hat{\Psi}_{\bf k}^{\dagger}\pmb{\cal S}\hat{\Psi}_{\bf k}$~\cite{offdiagonal_magnon_spin,spin_matrix}.
Under the magnon basis $\hat{\Psi}_{\bf k} = (\hat{b}_{i,\bf k},\hat{b}_{i,-\bf k}^\dagger)^T$, $\pmb{\cal S}$ is a $2N\times 2N$ matrix and represents the spin carried by magnon modes. In this matrix representation, the diagonal elements characterize the magnon spin carried by each mode, while the off-diagonal elements reflect superpositions of different modes. Nonzero diagonal terms correspond to circularly polarized modes, such as those in collinear easy-axis AFMs~\cite{AFM1,AFM2,AFM3,AFM4,AFM5} and altermagnets~\cite{ATM1,ATM2}, whereas their absence indicates linear polarization, as in easy-plane AFMs~\cite{AFM_plane1,AFM_plane2,AFM_plane3}. Referring to details in End Matter, the spin matrix hereby reads
\begin{align}
    {\cal S}_{x}({\bf k})=0,~~~~
    \mathcal{S}_{y,z}^{\xi,\xi'}({\bf k})\propto \sum_{j=1}^{N} e^{ij\left[\pm \phi + {2\pi}(\xi'-\xi)/{N}\right]}.
    \label{spin_matrix}
\end{align}
A nonzero matrix element requires $2\pi(\xi'-\xi)/N\pm\phi=2l\pi$ ($l\in\mathbb{Z}$), leading to the selection rule
\begin{align}
    \xi'-\xi=\pm m\pm Nl.
\label{selection_rule}
\end{align}
For example, when $N=5$ for a FM spiral, the selection rule \eqref{selection_rule} restricts the allowed magnon spins carried by $|\xi'-\xi|=\{1,4\}$, as illustrated in Fig.~\ref{fig3}. 
Crucially, the absence of all diagonal elements confirms that individual magnon modes in magnetic spirals carry zero spin, which is instead mediated entirely by mode coherence. Further, reversing the DMI flips the pitch angle $\phi \to -\phi$. 
The even parity of $\cos(j\phi)$ (see End Matter) renders the invariance of $\mathcal{S}_y$, while the odd parity of $\sin(j\phi)$ forces a inverted spin $\mathcal{S}_z^{\xi,\xi'} \to -\mathcal{S}_z^{\xi,\xi'}$ along the spiral, as addressed in Fig.~\ref{fig3}.

The selection rule \eqref{selection_rule} is governed by the magnetic screw symmetry $\hat{S}_{\rm scr}=\hat{R}_{x}(-\phi)\hat{T}_{z}$, which combines a one-site translation $\hat{T}_{z}$ along $\hat{\bf z}$ with a spin rotation $\hat{R}_{x}(-\phi)$ by $-\phi$ about $\hat{\bf x}$. Since $[\hat{H},\hat{S}_{\rm scr}]=0$, the eigenstates obey $\hat{S}_{\rm scr}\ket{\xi,{\bf k}}=\lambda\ket{\xi,{\bf k}}$ with eigenvalue $\lambda$. Applying $\hat{S}_{\rm scr}^N=\hat{R}_{x}(-N\phi)\hat{T}_{z}^N=\hat{T}_{z}^N$ over one magnetic period yields $\lambda^N=e^{ik_zN\delta}$, where $\delta$ is the lattice constant. Thereby, within a fixed momentum sector, $\hat{S}_{\rm scr}$ acts as
\begin{equation}
\hat{S}_{\rm scr}\ket{\xi,{\bf k}}=\exp\left[i\left(k_z\delta+{2\pi(\xi-1)}/{N}\right)\right]\,\ket{\xi,{\bf k}},
\label{screwcharge}
\end{equation}
where $k_z\delta$ is the translational phase shift and $2\pi(\xi-1)/N$ is the winding of the mode's pattern within the magnetic cell. Since the cycloid lies in the $yz$-plane, the magnon spin components are restricted to the $y$- and $z$-components.
Defining the ladder operators $\hat S_{\pm}=\sum_{i=1}^N(\hat S^y_i\pm i\hat S^z_i)$ associated with $U(1)$ spin rotations about $\hat{\bf x}$, their transformation under the screw operation reads $\hat{S}_{\rm scr}\hat{S}_{\pm}\hat{S}_{\rm scr}^{-1}=e^{\mp i\phi}\hat{S}_{\pm}$. Inserting $\hat{S}_{\rm scr}^{-1}\hat{S}_{\rm scr}$ into the matrix element and employing Eq.~\eqref{screwcharge} yields
\begin{equation}
\bra{\xi,{\bf k}}\hat S_{\pm}\ket{\xi^\prime,{\bf k}}
=e^{\,i{2\pi}(\xi^\prime-\xi)/N\mp i\phi}
\bra{\xi,{\bf k}}\hat S_{\pm}\ket{\xi^\prime,{\bf k}},
\label{insertion}
\end{equation}
which reproduces the selection rule in \eqref{selection_rule}. 
Further, the sign reversal of $D$ is equivalent to a mirror operation $\hat{M}_y$ with respect to the $x$-$z$-plane. Due to the axial-vector nature of spin,  $\hat{M}_y^{-1}\hat{S}_{z}\hat{M}_y=-\hat{S}_z$ and $\hat{M}_y^{-1}\hat{S}_{y}\hat{M}_y=\hat{S}_y$, forcing the magnon spin matrix elements to transform as $({\cal S}_y^{\xi,\xi'},{\cal S}_z^{\xi,\xi'})\rightarrow({\cal S}_y^{\xi,\xi'},-{\cal S}_z^{\xi,\xi'})$.

The vanished diagonal elements in $\pmb{\cal S}$ imply that all magnon modes are linearly polarized. The polarization of mode-$\xi$ is captured by the linear terms in the spin operator (see End Matter), governed by the vector $\delta{\bf S}_{\bf q}^{\xi}=\delta{S}_{{\bf q},x}^{\xi}\hat{\bf x}+i\delta{S}_{{\bf q},y}^{\xi}\hat{\bf y}+i\delta{S}_{{\bf q},z}^{\xi}\hat{\bf z}$. We find that the real and imaginary parts cannot coexist simultaneously, such that the magnon modes exhibit either linear polarization normal to or within the spiral $yz$-plane, with explicit wave-vector dependence. This reveals a profound physical implication: spin transport in \textit{non-canted} chiral spirals is mediated entirely by quantum coherence, implying that the spin transport is favorably driven via electrical injection rather than the spin Seebeck effect~\cite{spin_matrix,Bauer2012} that involves only magnon population. Magnetic canting enables the spin Seebeck effect with spin polarized along the canting direction~\cite{yamasaki_PRL_2007,Ramesh_PRL_2022,Ramesh_NM_2024,Ramesh_NC_2024,Ramesh_AM_2024,Ramesh_npj_2025,Ramesh_PRL_2025}.

\textit{Quantum spin transport}.---We then formulate the electrical spin injection across the normal-metal (NM)/magnet interface biased by electron spin accumulation $\Delta\mu\hat{\bf z}$ and magnon spin diffusion in the spiral magnet (Fig.~\ref{fig1}). The thickness of the NM is $d$, and $a$ is its lattice constant.
Magnons couple to electrons via the interfacial $s$-$d$ exchange interaction with the exchange coupling strength $J_{\rm sd}$~\cite{s-d_interaction_1,s-d_interaction_2,s-d_interaction_3,s-d_interaction_4,s-d_interaction_5}, governed by
\begin{align}
        \hat{H}_{\rm int}&=\frac{a}{Nd}\sqrt{\frac{S}{2N_c}}J_{\rm sd}\sum_{\xi=1}^N\sum_{\alpha,\beta=\uparrow,\downarrow}\sum_{{\bf k},{\bf k}',\bf q}\left({\pmb \sigma}_{\alpha\beta}\cdot \delta{\bf S}_{{\bf q}}^{\xi}\right)\nonumber\\
        &\times\hat{c}_{{\bf k}'\alpha}^{\dagger}\hat{c}_{{\bf k}\beta}\hat{b}_{\xi,{\bf q}}\delta_{{\bf k}+{\bf q},{\bf k}'}+\text{H.c.},
        \label{H_int}
\end{align}
where $N_c$ is the total number of unit cells in the magnet, $\pmb{\sigma}$ is the vector of Pauli matrices, and $\hat{c}_{\mathbf{k}\beta}$ is the electron annihilation operator with spin $\beta$.
We establish a quantum kinetic equation dealing with spin injection, diffusion, and relaxation on equal footing for the magnon density matrix $\rho_{\bf q}=\langle\hat{\Psi}_{\bf q}^{\dagger T}\hat{\Psi}_{\bf q}^T\rangle$ (see End Matter), which characterizes all procedures of non-local transport mediated by magnon spins. 
Here, with $\langle\cdots\rangle$ denoting the ensemble average, the diagonal terms of $\rho_{\bf q}$ represent magnon population in mode-$\xi$ and the off-diagonal terms represent quantum coherence~\cite{Yu2013}.

\begin{figure}[htp!]
    \centering
    \includegraphics[width=9.0cm]{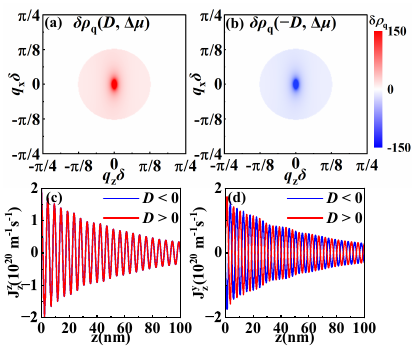}
    \caption{Injected magnon quantum coherence $\delta\rho_{\bf q}$ and spin currents with opposite structural chirality, calculated using parameters $N=5$, $K=0.35~{\rm meV}$, $\phi=2\pi/N$, $J=-K\cos\phi$, $|D|=K\sin{\phi}$, $J_{\rm sd}=100~{\rm meV}$, $d=4~{\rm nm}$, and $\Delta\mu=0.2~{\rm meV}$. (a) and (b) plot the injected quantum coherence $\rho({\bf q})|_{15}$. (c) and (d) plot injected magnon spin currents $J_z^z$ polarized along $\hat{\bf z}$ and $J_z^y$ polarized along $\hat{\bf y}$ as a function of diffusion distance $z$ for opposite chirality $D<0$ (blue curves) and $D>0$ (red curves).}
    \label{fig4}
\end{figure}

The spin injection/detection efficiency is then governed by the scattering amplitude~\cite{supplement} 
\begin{align}
\sum_{\alpha,\beta=\uparrow,\downarrow}{\cal Z}^{\xi\xi'}_{\alpha\beta}({\bf q})=2\Delta\mu\left(\delta{S}_{{\bf q},y}^{\xi}\delta{S}_{{\bf q},x}^{\xi'*}-\delta{S}_{{\bf q},x}^{\xi}\delta{S}_{{\bf q},y}^{\xi'*}\right)
\end{align}
between modes $\xi$ and $\xi'$ mediated by $\Delta\mu\hat{\bf z}$.
Since the $y$-component of the polarization $\delta{S}_{{\bf q},y}^{\xi}(D)=-\delta{S}_{{\bf q},y}^{\xi}(-D)$ is odd under inversion of $D$, whereas the $x$-component is even [see End Matter Eq.~\eqref{deltaS}], ${\cal Z}^{\xi\xi'}_{\alpha\beta}({\bf q})$ and all the injected magnon quantum coherence $\rho_{\bf q}|_{\xi\ne \xi'}$ are odd under chirality reversal: $\delta\rho_{\mathbf{q}}(D, \Delta\boldsymbol{\mu})=-\delta\rho_{\mathbf{q}}(-D, \Delta\boldsymbol{\mu})$. Taking $N=5$ in the FM spiral as an example: the reversal of ${\rm Re}[\rho_{15}]$ shown in Fig.~\ref{fig4}(a) and (b) demonstrates the locking of electrically injected magnon quantum coherence with structural chirality.

The electrically-injected magnon quantum coherence at $z=0$ diffuse along $\hat{\bf z}$ in the magnet of area $S_0$, resulting in the spin currents polarized along $\alpha=\{y,z\}$ at $z>0$ (see details in the SM~\cite{supplement})
\begin{align}
    J^{\alpha}_{z}(z)&=\frac{2}{S_0}\sum_{\bf q}\sum_{\xi,\xi'=1}^{N}{\rm Re}\Big[v_{\xi\xi'}^{z}({\bf q}){\cal S}_{\alpha}^{\xi,\xi'}({\bf q})\rho_{\xi,\xi'}({\bf q})\nonumber\\
        &\times \exp\left({-\frac{z}{{v}_{\xi\xi'}^{z}({\bf q})\tau_{\rm eff}}}\right)
    +{\tilde v}_{\xi\xi'}^{z}({\bf q}){\cal S}_{\alpha}^{\xi,\xi'+N}({\bf q})\rho_{\xi,\xi'+N}({\bf q})\nonumber\\
    &\times \exp\left({-\frac{z}{{\tilde v}_{\xi\xi'}^{z}({\bf q}){\tilde \tau}_{\rm eff}}}\right)\Big],
    \label{spin_current}
\end{align}
noting ${\cal S}_{\alpha}^{\xi,\xi'}({\bf q})$ is the magnon spin carried by mode-$(\xi,\xi')$ superposition. The mode-averaged group velocities along diffusion $\hat{\bf z}$ read $v_{\xi\xi'}^{z}({{\bf q}})=[v_{\xi}^{z}({\bf q})+v_{\xi'}^{z}({\bf q})]/2$ and ${\tilde v}_{\xi\xi'}^{z}({\bf q})=[v_{\xi}^{z}({\bf q})+v_{\xi'}^{z}(-{\bf q})]/2$; $1/\tau_{\rm eff}=1/\tau_m-i\left(\omega_{\xi}({\bf q})-\omega_{\xi'}({\bf q})\right)$ and $1/{\tilde \tau}_{\rm eff}=1/\tau_m-i\left(\omega_{\xi}({\bf q})+\omega_{\xi'}(-{\bf q})\right)$ are effective decay rates. Both spin currents polarized along $\hat{\bf y}$ and $\hat{\bf z}$ exhibit quantum beats during diffusion [Fig.~\ref{fig4}(c) and (d)], originating from the frequency differences between different bands. Crucially, their dependence on structural chirality differs: the $z$-polarized component $J_z^z$ is even under chirality reversal [Fig.~\ref{fig4}(c)], whereas the $y$-polarized component $J_z^y$ is odd [Fig.~\ref{fig4}(d)]. This parity behavior stems from 
\begin{align}
    \mathrm{Parity}[J_{z}^{\alpha}(D)]=\mathrm{Parity}\left[{\cal S}_{\alpha}(D)\delta\rho_{\mathbf q}(D)\right].
    \label{parity}
\end{align}
Since $D\to-D$ makes ${\cal S}_z$ and $\delta\rho_{\mathbf q}$ odd but ${\cal S}_y$ even, $J_{z}^{z}$ is even and $J_{z}^{y}$ is odd of the structural chirality.

Reversed voltage in a non-local setup~\cite{nonlocal} relates to the magnon spin accumulation polarized along $\hat{\bf z}$ or magnon spin current $J_z^z$ (see below), while $J_z^y$ and injected magnon population contribute to $D$-even voltage. According to relation \eqref{parity}, the injected spin current $J_z^z$ is invariant under chirality reversal. 
However, this does not imply that the non-local voltage is not reversed by the structural chirality when the detection process is accounted for: the spin-conversion process at the detector is also inverted by $D$. Indeed, at the detection interface, the magnon accumulation $\delta\rho_{\bf q}$ induces a pure electron spin current back into the metal contact, estimated by~\cite{s-d_interaction_1}
\begin{align}
&j_y^z=-(1/S_0)d\langle\hat{S}_{z}\rangle/dt\nonumber\\
 &=-\frac{{Sa^4m^2J_{\rm sd}^2}}{16\pi{S}_0dq\hbar^5}\sum_{\bf q}\sum_{\xi,\xi'=1}^N\delta\rho_{\bf q}{\cal S}_{z}^{\xi,\xi'}\tilde{\cal X}^{\xi\xi'}({\bf q})+\rm{H.c.},
    \nonumber
\end{align}
where ${\cal X}^{\xi\xi'}({\bf q})=-2\hbar\omega_{\xi,{\bf q}}\left(\delta{S}_{{\bf q},y}^{\xi}\delta{S}_{{\bf q},x}^{\xi'*}-\delta{S}_{{\bf q},x}^{\xi}\delta{S}_{{\bf q},y}^{\xi'*}\right)$
is odd in $D$. Namely, the resulting electron spin current
\begin{align}
    &\mathrm{Parity}[j_{y}^{z}(D)]=\mathrm{Parity}\left[{\cal S}_{z}(D)\delta\rho_{\mathbf q}(D){\cal X}(D)\right]
\end{align}
is odd in $D$ since all three terms are odd in it. This yields a reversed voltage response $V(D)=-V(-D)$ via the inverse spin Hall effect $V\propto j_y\hat{\bf y}\otimes \sigma_z\hat{\bf z}$. Thereby, a switched non-local voltage in the first harmonic~\cite{yamasaki_PRL_2007,Ramesh_PRL_2022,Ramesh_NM_2024,Ramesh_NC_2024,Ramesh_AM_2024,Ramesh_npj_2025} by an electric field is traced to the interface effect rather than bulk transport.

\textit{Conclusion and discussion}.---In summary, we show in electrically-driven magnon transport of both FM and AFM spirals that the structural chirality reverses the non-local voltage mediated by magnons without changing the spin current that produces it: screw symmetry confines magnon spin to inter-band quantum coherence, making chirality a property of the detector rather than of transport.  Two consequences follow: the quantum coherence imprints damped spatial beats on the diffusive spin current, set by inter-band frequency differences; a thermal spin-Seebeck drive, populating modes without coherence, cannot access spin transport unless canting opens a separate channel. These conclusions do not rely on commensurability: for an incommensurate spiral the discrete sum is replaced by an average over one spiral period, so the screw symmetry still forbids a diagonal spin matrix element and chirality-odd readout persists as $N\rightarrow \infty$. They should therefore extend to other spiral insulators with electrically switchable DMI. Structural chirality thus emerges as an intrinsic control parameter for spin-to-charge conversion at the interface, providing a route to chirality-controlled (quantum) information processing in spintronics.

\begin{acknowledgments}
This work is financially supported by the National Natural Science Foundation of China under Grant No.~12374109 and the National Key Research and Development Program of China under Grant No.~2023YFA1406600.  J.Z. acknowledges the Quantum Center at KFUPM for the support received under No.~INQC2600. S.S. acknowledges the support of a fellowship from ``la Caixa" Foundation (ID 100010434) with the fellowship code LCF/BQ/PI25/12100023, and grant ``Excelencia Mar\'ia de Maeztu" CEX2023-001316-M funded by MICIU/AEI /10.13039/501100011033.
\end{acknowledgments}

\section{End Matter}

\textit{Magnon spin matrix}.---The Hamiltonian matrix ${\cal H}_{\bf k}={\rm diag}({\cal H}^{(1)}_{{\bf k}},{\cal H}^{(2)}_{{\bf k}},\cdots,{\cal H}^{(N)}_{{\bf k}})$ for the HP bosons, as addressed in the main text, is block diagonalized with elements ($\xi=\{1,\cdots,N\}$)~\cite{supplement}
\begin{align}
    {\cal H}^{(\xi)}_{{\bf k}}=\left(\begin{array}{cc}
      A_\xi({\bf k}) & B_\xi({\bf k}) \\
      B_\xi({\bf k}) & A_\xi({\bf k})
    \end{array}\right),
\end{align}
in which $A_\xi({\bf k})$ and $B_\xi({\bf k})$ for the FM spiral are explicitly given in the main text. For the AFM spiral, the coefficients~\cite{supplement}
\begin{align}
    A_\xi({\bf k})&=-KS(\cos\phi+1)\cos(k_z\delta+2\pi (\xi-1)/N)\nonumber\\
    &+2KS(1-\cos\phi),\nonumber\\
    B_\xi({\bf k})&=-2KS\cos\phi\cos(k_x\delta)-KS(\cos\phi-1)\nonumber\\
    &\times\cos(k_z\delta+2\pi (\xi-1)/N),
\end{align}
are also even functions of the DMI parameter $D$.
To diagonalize the magnon Hamiltonian ${\cal H}^{(\xi)}_{{\bf k}}$ and obtain the magnon dispersion, we apply the $2\times 2$ Bogoliubov transformation
\begin{align}
{\cal T}^{(\xi)}_{\bf k}=\left(\begin{array}{cc}
\cosh{\gamma_{\xi}}&\sinh{\gamma_{\xi}}\\
     \sinh{\gamma_{\xi}}&\cosh{\gamma_{\xi}}
    \end{array}\right),
    \label{B_matrix}
\end{align}
in which $\cosh{\gamma_{\xi}}=[(A_{\xi}({\bf k})+\hbar\omega_{\xi,{\bf k}})/(2\hbar\omega_{\xi,{\bf k}})]^{1/2}$, $\sinh{\gamma_\xi}=[(A_\xi({\bf k})-\hbar\omega_{\xi,{\bf k}})/(2\hbar\omega_{\xi,{\bf k}})]^{1/2}$, and $\tanh({2\gamma_{\xi}})=-B_\xi({\bf k})/A_\xi({\bf k})$. 
Through this transformation
\begin{align}
   ({\cal T}^{(\xi)}_{\bf k})^{-1}(\tilde{\sigma}_3{\cal H}^{(\xi)}_{{\bf k}}){\cal T}^{(\xi)}_{\bf k}&=\text{diag}\left(\hbar\omega_{\xi,\bf k},-\hbar\omega_{\xi,-\bf k}\right),
\end{align}
where $\tilde{\sigma}_3=\text{diag}(1,-1)$ serves as the metric, 
we obtain the magnon dispersion $\hbar\omega_{\xi,{\bf k}}$. Accordingly, the full Bogoliubov transformation $2N\times 2N$ matrix ${\cal T}({\bf k})$ contains a proper arrange of Eq.~\eqref{B_matrix}~\cite{supplement}, with the elements given by ($j=\{1,2,\cdots,N\},~\xi=\{1,\cdots,N\}$)
\begin{align}
    &{\cal T}_{j,\xi}({\bf k})=\frac{1}{\sqrt{N}}{\rm exp}\big[i2{\pi}(\xi-1)(j-1)/N\big]\cosh{\gamma_{\xi}},\nonumber\\
    &{\cal T}_{j,\xi+N}({\bf k})=\frac{1}{\sqrt{N}}{\rm exp}\big[i2{\pi}(\xi-1)(j-1)/N\big]\sinh{\gamma_{\xi}}.
    \label{t_k}
\end{align}

The matrix elements \eqref{t_k} help to find the spin matrix $\pmb{\cal S}({\bf k})$ under the magnon basis $\hat{\Psi}_{\bf k} = (\hat{b}_{\xi,\bf k},\hat{b}_{\xi,-\bf k}^\dagger)^T$, which appears in the spin operator 
\begin{align}
\hat{\bf S}=(1/2)\sum_{\bf k}\hat{\Psi}_{\bf k}^{\dagger}\pmb{\cal S}({\bf k})\hat{\Psi}_{\bf k}+{\rm linear~~terms}.
\label{spin_operator}
\end{align}
With details given in the SM~\cite{supplement}, the $2N\times2N$ magnon spin matrix ${\pmb {\cal S}}$ has components
\begin{align}
    {\cal S}_{x}({\bf k})&=0,\nonumber\\
    {\cal S}_{y}({\bf k})&=-\sum_{j=1}^N\cos{(j\phi)}
    \left(
    \begin{array}{cc}
     {\cal A}_j({\bf k})|_{{N}\times{N}}    &  {\cal B}_j({\bf k})|_{{N}\times{N}} \\
    {\cal B}_j({-\bf k})^*|_{{N}\times{N}}     & {\cal A}_j({-\bf k})^*|_{{N}\times{N}}
    \end{array}
    \right),\nonumber\\
   {\cal S}_{z}({\bf k})&=\sum_{j=1}^N\sin{(j\phi)}
    \left(
    \begin{array}{cc}
     {\cal A}_j({\bf k})|_{{N}\times{N}}    &  {\cal B}_j({\bf k})|_{{N}\times{N}} \\
    {\cal B}_j({-\bf k})^*|_{{N}\times{N}}     & {\cal A}_j({-\bf k})^*|_{{N}\times{N}}
    \end{array}
    \label{spin matrix}
    \right),
\end{align}
viz., the magnons carry spins only along the $\hat{\bf y}$- and $\hat{\bf z}$-directions. Here, the $(\xi,\xi')$-th element of ${\cal A}_j({\bf k})$ and ${\cal B}_j({\bf k})$ reads
\begin{align}
    {\cal A}_j|_{\xi, \xi'}&=\frac{1}{N} \big({\rm exp}\big[{i2\pi j(\xi'-\xi)/N}\big]\cosh\gamma_\xi\cosh\gamma_{\xi'}\nonumber\\
    &+{\rm exp}\big[{-i2\pi j(\xi'-\xi)/N}\big]\sinh\gamma_\xi \sinh\gamma_{\xi'}\big),\nonumber\\
    {\cal B}_j|_{\xi, \xi'}&=\frac{1}{N} \big({\rm exp}\big[{i2\pi j(\xi'-\xi)/N}\big]\cosh\gamma_\xi\sinh\gamma_{\xi'}\nonumber\\
    &+{\rm exp}\big[{-i2\pi j(\xi'-\xi)/N}\big]\cosh\gamma_\xi \sinh\gamma_{\xi'}\big).
\end{align}
In the main text, these expressions are addressed in Eq.~\eqref{spin_matrix} for both the FM and AFM spirals. The nonzero elements of ${\cal A}_{j}({\bf k})$ and ${\cal B}_{j}({\bf k})$ produce the selection rule Eq.~\eqref{selection_rule} in the main text. This selection rule applies with $m=1$ for FM, but with other choices of $N/4<m<N/2$ and ${\rm gcd}(m,N)=1$ for the AFM spiral~\cite{supplement}.

\textit{Linear polarizations of spin-wave modes}.---Beyond the quadratic terms, the linear terms in the spin operator $\hat{\bf S}$ \eqref{spin_operator} represent the polarization of magnon modes~\cite{spin_matrix}, which is characterized by the polarization vector $\delta{\bf S}_{\bf q}^{\xi}$ for mode $\xi$. The components of this vector read~\cite{supplement} 
\begin{align}
    \delta{S}_{{\bf q},x}^{\xi}&=\sum_{j=1}^{N}\delta S_{{\bf q},x}^{\xi,j}=\sum_{j=1}^{N}\left({\cal T}_{j,\xi}({\bf q})+{\cal T}_{j,\xi+N}^{*}({-\bf q})\right),\nonumber\\
    \delta{S}_{{\bf q},y}^{\xi}&=\sum_{j=1}^{N}\delta S_{{\bf q},y}^{\xi,j}=\sum_{j=1}^{N}\left({\cal T}_{j,\xi}({\bf q})-{\cal T}_{j,\xi+N}^{*}({-\bf q})\right)\sin(j\phi),\nonumber\\
    \delta{S}_{{\bf q},z}^{\xi}&=\sum_{j=1}^{N}\delta S_{{\bf q},z}^{\xi,j}=\sum_{j=1}^{N}\left({\cal T}_{j,\xi}({\bf q})-{\cal T}_{j,\xi+N}^{*}({-\bf q})\right)\cos({j\phi}).
    \label{deltaS}
\end{align}
According to Eq.~\eqref{t_k}, the elements of matrix ${\cal T}_{\bf k}$ are determined by the coefficients $A_{\xi}({\bf k})$ and $B_{\xi}({\bf k})$. Since all of them are invariant under the reversal of the structural chirality, the matrix ${\cal T}_{\bf k}$ remains invariant. Consequently, the even parity of $\cos(j\phi)$ guarantees the invariance of $\delta{S}_{{\bf q},z}^{\xi}$ under the reversal of  structural chirality $D$, while the odd parity of $\sin(j\phi)$ enforces an exact sign inversion 
for $\delta{S}_{{\bf q},y}^{\xi}$.

\textit{Quantum kinetic equation}.---The interfacial coupling between magnetic moments in the chiral spiral magnet and electrons in the NM with thickness $d$ and lattice constant $a$ is mediated by the $s$-$d$ exchange interaction, governed by the Hamiltonian
\begin{align}
    \hat{H}_{\rm int}=J_{\rm sd}a^3\sum_{l=1}^{N_c}\sum_{j=1}^{N}\int d{\bf r}\delta({\bf r}-{\bf R}_{l,j})\hat{\bf s}({\bf r})\cdot\hat{\bf S}_{l,j},
\end{align}
where $\hat{\bf s}({\bf r})$ is the electron spin density, $N_c$ is the total number of unit cells in the magnet, and $N$ is the number of sites in one magnetic unit cell. Taking into account the electric spin injection, we establish a quantum kinetic equation to find the non-equilibrium evolution of the magnon density matrix 
\begin{align}
    {\rho}_{\bf q}=\left(
    \begin{array}{c|c}
    \langle\hat{b}_{\xi,\bf q}^{\dagger}\hat{b}_{\xi',\bf q}\rangle|_{N\times N} & \langle\hat{b}_{\xi,{\bf q}}^{\dagger}\hat{b}_{\xi',-{\bf q}}^{\dagger}\rangle|_{N\times N}\\
    \hline
    \langle\hat{b}_{\xi,-\bf q}\hat{b}_{\xi',\bf q}\rangle|_{N\times N} & \langle\hat{b}_{\xi,-\bf q}\hat{b}_{\xi',-{\bf q}}^{\dagger}\rangle|_{N\times N}
    \end{array}
    \right),
\end{align}
in which $\{\xi,\xi'\}=\{1,\cdots,N\}$ represent the magnon modes, which obeys the quantum kinetic equation~\cite{Yu2013}
\begin{align}
    \partial_t\rho_{\bf q}=\partial_t\rho_{\bf q}|_{\rm{inj}}+\partial_t\rho_{\bf q}|_{\rm{coh}}+\partial_t\rho_{\bf q}|_{\rm{rex}}.
\end{align}

Driven by the electron spin accumulation $\Delta\mu=\delta\mu_{\uparrow}-\delta\mu_{\downarrow}$ along $\hat{\bf z}$, the  electric injection of multiband magnons is governed by
\begin{align}
    \partial_{t}\rho_{\bf q}|_{\text{inj}}&=\frac{\pi}{\hbar}\frac{S}{2NN_c}\frac{a^2}{d^2}J_{\rm sd}^2\sum_{\bf k}\sum_{\alpha,\beta=\uparrow,\downarrow}\big[f_{{\bf k+q},\alpha}\left(1-f_{{\bf k},\beta}\right)\sigma_3\nonumber\\
    &+\rho_{\bf q}\left(f_{{\bf k+q},\alpha}-f_{{\bf k},\beta}\right)\big]{\cal M}_{\alpha\beta}({\bf k},{\bf k+q})+\text{H.c.},
    \label{quantum kinetic equation}
\end{align}
where $S=|{\bf S}_{l,i}|$ is the magnitude of magnetic moments in the magnet, $f_{{\bf k},\beta}=[f_{\bf k}^{(0)}-(\partial{f}_{\bf k}^{(0)}/\partial\varepsilon_{\bf k})\delta\mu_{\beta}]$ describes the non-equilibrium population of electrons with spin-$\beta$ that is shifted from the equilibrium Fermi-Dirac distribution $f_{\bf k}^{(0)}$, in which $\varepsilon_{\bf k}$ is the electron kinetic energy, and the $2N\times2N$ matrix ${\cal M}_{\alpha\beta}({\bf k},{{\bf k}+{\bf q}})$ shares a similar form with the scattering cross section ${\cal Z}^{\xi\xi'}_{\alpha\beta}({\bf q})$ in the main text, also given in a detail in the SM~\cite{supplement}.

The coherent dynamics of the magnon density matrix is governed by the magnon dispersion matrix ${\cal E}_{\bf q}=\text{diag}\left(\hbar\omega_{1,{\bf q}},\cdots,\hbar\omega_{N,{\bf q}},\hbar\omega_{1,-{\bf q}},\cdots,\hbar\omega_{N,-{\bf q}}\right)$ via 
\begin{align}
    \partial_{t}\rho_{\bf q}|_{\text{coh}}=\frac{1}{i\hbar}\big(\rho_{\bf q}\sigma_3{\cal E}_{\bf q}-{\cal E}_{\bf q}\sigma_3\rho_{\bf q}\big).
\end{align}
Furthermore, the relaxation toward the thermal equilibrium Bose-Einstein distribution $\rho_{\bf q}^{(0)}$ of magnon can be well parameterized within the relaxation-time approximation~\cite{relaxation-time1,relaxation-time2} as 
\begin{align}
    \partial_t\rho_{\bf q}|_{\rm rex}=-(\rho_{\bf q}-\rho_{\bf q}^{(0)})/\tau_m,
\end{align}
in which we assume a mode-independent relaxation time $\tau_m$ for simplicity.

The electrically injected magnon below one NM electrode at $z=0$ diffuses towards the other contact and induces an electron spin current and non-local voltage signals there. The spatial diffusion of the magnon distribution and coherence in the magnet is governed by
\begin{align}
    &-v_{\xi\xi'}^z({\bf q})\frac{\partial\rho_{\xi,\xi'}(z,{\bf q})}{\partial z}-\frac{1}{\tau_{\rm eff}}\rho_{\xi,\xi'}(z,{\bf q})=0,\nonumber\\
    &-{\tilde v}_{\xi\xi'}^z({\bf q})\frac{\partial\rho_{\xi,\xi'+N}(z,{\bf q})}{\partial z}-\frac{1}{\tilde \tau_{\rm eff}}\rho_{\xi,\xi'+N}(z,{\bf q})=0.
    \label{diffusion_equation}
\end{align}
Here, 
\begin{align}
    v_{\xi\xi'}^{z}({{\bf q}})=\frac{v_{\xi}^{z}({\bf q})+v_{\xi'}^{z}({\bf q})}{2},~~~~{\tilde v}_{\xi\xi'}^{z}({\bf q})=\frac{v_{\xi}^{z}({\bf q})+v_{\xi'}^{z}(-{\bf q})}{2},
\end{align}
are the mode-averaged group velocities along the transport $\hat{\bf z}$-direction; 
\begin{align}
    &1/\tau_{\rm eff}=1/\tau_m-i\left(\omega_{\xi}({\bf q})-\omega_{\xi'}({\bf q})\right),\nonumber\\
    &1/{\tilde \tau}_{\rm eff}=1/\tau_m-i\left(\omega_{\xi}({\bf q})+\omega_{\xi'}(-{\bf q})\right),
\end{align}
are the effective relaxation rates of the non-equilibrium magnon coherence.


\begin{thebibliography}{99}

%--------chiral_phonon----------

\bibitem{chiral_phonon1} L. Zhang and Q. Niu, Chiral phonons at high-symmetry points in monolayer hexagonal lattices, Phys. Rev. Lett. \textbf{115}, 115502 (2015).

\bibitem{chiral_phonon2} H. Zhu, J. Yi, M.-Y. Li, J. Xiao, L. Zhang, C.-W. Yang, R. A. Kaindl, L.-J. Li, Y. Wang, and X. Zhang, Observation of chiral phonons, Science \textbf{359}, 579 (2018).

\bibitem{chiral_phonon3} K. Ishito, H. Mao, Y. Kousaka, Y. Togawa, S. Iwasaki, T. Zhang, S. Murakami, J.-i. Kishine, and T. Satoh, Truly chiral phonons in $\alpha$-HgS, Nat. Phys. \textbf{19}, 35 (2023).

\bibitem{chiral_phonon4} H. Ueda, M. Garc\'ia-Fern\'andez, S. Agrestini, C. P. Romao, J. van den Brink, N. A. Spaldin, K.-J. Zhou, and U. Staub, Chiral phonons in quartz probed by X-rays, Nature \textbf{618}, 946 (2023).

%---------------chiral_magnon------------------

\bibitem{chiral_magnon1} R. W. Damon and J. R. Eshbach, Magnetostatic modes of a ferromagnet slab, J. Phys. Chem. Solids \textbf{19}, 308 (1961).

\bibitem{chiral_magnon2} R. Cheng, S. Okamoto, and D. Xiao, Spin Nernst Effect of Magnons in Collinear Antiferromagnets, Phys. Rev. Lett. \textbf{117}, 217202 (2016).

\bibitem{chiral_magnon3} L. \v{S}mejkal, A. Marmodoro, K.-H. Ahn, R. Gonz\'alez-Hern\'andez, I. Turek, S. Mankovsky, H. Ebert, S. W. D'Souza, O. \v{S}ipr, J. Sinova, and T. Jungwirth, Chiral magnons in altermagnetic RuO$_2$, Phys. Rev. Lett. \textbf{131}, 256703 (2023).



\bibitem{qing2026} Y. Qing, J. Zou, Y.-P. Wang, G. E. W. Bauer, and T. Yu, Universal Right-Hand Chirality of Evanescent Vector Fields, arXiv:2606.01585.

\bibitem{puebla2026} J. Puebla, J.-i. Kishine, M. Shiraishi, and K. Y. Bliokh, Chirality in Condensed Matter: Symmetry, Waves, and Quasiparticles, arXiv:2609.12434.


%-------wave-field--------------

\bibitem{wave1} K. Y. Bliokh, A. Y. Bekshaev, and F. Nori, Extraordinary momentum and spin in evanescent waves, Nat. Commun. \textbf{5}, 3300 (2014).


\bibitem{wave2} J. Petersen, J. Volz, and A. Rauschenbeutel, Chiral nanophotonic waveguide interface based on spin-orbit interaction of light, Science \textbf{346}, 67 (2014).


\bibitem{wave3} K. Y. Bliokh, D. Smirnova, and F. Nori, Quantum spin Hall effect of light, Science \textbf{348}, 1448 (2015).


\bibitem{wave4} P. Lodahl, S. Mahmoodian, S. Stobbe, A. Rauschenbeutel, P. Schneeweiss, J. Volz, H. Pichler, and P. Zoller, Chiral quantum optics, Nature \textbf{541}, 473 (2017).


\bibitem{wave5} S. Luo, L. He, and M. Li, Spin-momentum locked interaction between guided photons and surface electrons in topological insulators, Nat. Commun. \textbf{8}, 2141 (2017).


\bibitem{wave6} Q. Guo, T. Fu, J. Tang, D. Pan, S. Zhang, and H. Xu, Routing a chiral Raman signal based on spin-orbit interaction of light, Phys. Rev. Lett. \textbf{123}, 183903 (2019).

\bibitem{wave7} T. Yu, X. Zhang, S. Sharma, Y. M. Blanter, and G. E. W. Bauer, Chiral coupling of magnons in waveguides, Phys. Rev. B \textbf{101}, 094414 (2020).

\bibitem{wave8} Y. Yao, X. Li, and G. Chai, Unidirectional edge spin waves induced by spin-momentum locking of elastic waves in magnetic waveguides, Appl. Phys. Lett. \textbf{128}, 202407 (2026).


%----nonreciprocal transport-----

\bibitem{transport1} F. J. Rodr\'iguez-Fortu\~no, G. Marino, P. Ginzburg, D. O'Connor, A. Mart\'inez, G. A. Wurtz, and A. V. Zayats, Near-field interference for the unidirectional excitation of electromagnetic guided modes, Science \textbf{340}, 328 (2013).


\bibitem{transport2} R. J. Coles, D. M. Price, J. E. Dixon, B. Royall, E. Clarke, P. Kok, M. S. Skolnick, A. M. Fox, and M. N. Makhonin, Chirality of nanophotonic waveguide with embedded quantum emitter for unidirectional spin transfer, Nat. Commun. \textbf{7}, 11183 (2016).

\bibitem{transport3} T. Yu, Z. Luo, and G. E. Bauer, Chirality as generalized spin–orbit interaction in spintronics, Physics Reports \textbf{1009}, 1 (2023).


\bibitem{transport4} J. Zou, S. Bosco, E. Thingstad, J. Klinovaja, and D. Loss, Dissipative spin-wave diode and nonreciprocal magnonic amplifier, Phys. Rev. Lett. \textbf{132}, 036701 (2024).



%----------CISS----------------

\bibitem{CISS1} K. Ray, S. P. Ananthavel, D. H. Waldeck, and R. Naaman, Asymmetric scattering of polarized electrons by organized organic films of chiral molecules, Science \textbf{283}, 814 (1999).

\bibitem{CISS2} B. G\"ohler, V. Hamelbeck, T. Z. Markus, M. Kettner, G. F. Hanne, Z. Vager, R. Naaman, and H. Zacharias, Spin selectivity in electron transmission through self-assembled monolayers of double-stranded DNA, Science \textbf{331}, 894 (2011).

\bibitem{CISS3} R. Naaman, Y. Paltiel, and D. H. Waldeck, Chiral molecules and the electron spin, Nat. Rev. Chem. \textbf{3}, 250 (2019).

\bibitem{CISS4} A. Ghazaryan, Y. Paltiel, and M. Lemeshko, Analytic model of chiral-induced spin selectivity, J. Phys. Chem. C \textbf{124}, 11716 (2020).

\bibitem{CISS5} R. Naaman, Y. Paltiel, and D. H. Waldeck, Chiral molecules and the spin selectivity effect, J. Phys. Chem. Lett. \textbf{11}, 3660 (2020).


\bibitem{CISS6} B. Yan, Structural chirality and electronic chirality in quantum materials, Annu. Rev. Mater. Res. \textbf{54}, 97 (2024).



%----------skyrmion---------

\bibitem{sky1} S. M\"uhlbauer, B. Binz, F. Jonietz, C. Pfleiderer, A. Rosch, A. Neubauer, R. Georgii, and P. Boni, Skyrmion lattice in a chiral magnet, Science \textbf{323}, 915 (2009).


\bibitem{sky2} N. Nagaosa and Y. Tokura, Topological properties and dynamics of magnetic skyrmions, Nat. Nanotechnol. \textbf{8}, 899 (2013).


\bibitem{sky3} J. Barker and O. A. Tretiakov, Static and dynamical properties of antiferromagnetic skyrmions in the presence of applied current and temperature, Phys. Rev. Lett. \textbf{116}, 147203 (2016).


\bibitem{sky4} W. Jiang, X. Zhang, G. Yu, W. Zhang, X. Wang, M. B. Jungfleisch, J. E. Pearson, X. Cheng, O. Heinonen, K. L. Wang, Y. Zhou, A. Hoffmann, and S. G. E. te Velthuis, Direct observation of the skyrmion Hall effect, Nat. Phys. \textbf{13}, 162 (2017).



\bibitem{sky5} W. Legrand, D. Maccariello, F. Ajejas, S. Collin, A. Vecchiola, K. Bouzehouane, N. Reyren, V. Cros, and A. Fert, Room-temperature stabilization of antiferromagnetic skyrmions in synthetic antiferromagnets, Nat. Mater. \textbf{19}, 34 (2020).

%---------spiral---------------

\bibitem{spiral1} G. Shirane, R. Cowley, C. Majkrzak, J. B. Sokoloff, B. Pagonis, C. H. Perry, and Y. Ishikawa, Spiral magnetic correlation in cubic MnSi, Phys. Rev. B \textbf{28}, 4105 (1983).


\bibitem{spiral2} A. Rold\'{a}n-Molina, M. J. Santander, \'{A}. S. N\'{u}\~{n}ez, and J. Fern\'{a}ndez-Rossier, Quantum theory of spin waves in finite chiral spin chains, Phys. Rev. B \textbf{89}, 054403 (2014).


\bibitem{spiral3} O. I. Utesov, A. V. Sizanov, and A. V. Syromyatnikov, Spiral magnets with Dzyaloshinskii-Moriya interaction containing defect bonds, Phys. Rev. B \textbf{92}, 125110 (2015).


\bibitem{spiral4} N. Arakawa and J.-i. Ohe, Inplane anisotropy of longitudinal thermal conductivities and weak localization of magnons in a disordered spiral magnet, Phys. Rev. B \textbf{98}, 014421 (2018).


\bibitem{spiral5} F. J. d. Santos, M. d. S. Dias, and S. Lounis, Modeling spin waves in noncollinear antiferromagnets: Spin-flop states, spin spirals, skyrmions, and antiskyrmions, Phys. Rev. B \textbf{102}, 104436 (2020).



\bibitem{spiral6} H. W. Park, S. Zhang, P. Meisenheimer, M. Ramesh, S. Husain, I. Harris, J. \'{I}\~niguez-Gonz\'alez, Z. Yao, R. Ramesh, and S. K. Kim, Magnon thermal conductivity in multiferroics with spin cycloids, Phys. Rev. B \textbf{112}, 014414 (2025).

%----skyrmion Hall effct-------

\bibitem{denisov_skyrmion_2016} K. S. Denisov, I. V. Rozhansky, N. S. Averkiev, and E. L\"ahderanta, Electron scattering on a magnetic skyrmion in the nonadiabatic approximation, Phys. Rev. Lett. \textbf{117}, 027202 (2016).


\bibitem{iwasaki_magnon_skyrmion_2014} J. Iwasaki, A. J. Beekman, and N. Nagaosa, Theory of magnon-skyrmion scattering in chiral magnets, Phys. Rev. B \textbf{89}, 064412 (2014).


\bibitem{schuette_magnon_skyrmion_2014} C. Sch\"utte and M. Garst, Magnon-skyrmion scattering in chiral magnets, Phys. Rev. B \textbf{90}, 094423 (2014).


%---DMI------

\bibitem{DMI1} I. E. Dzyaloshinskii, Thermodynamic theory of weak ferromagnetism in antiferromagnetic substances, Sov. Phys. JETP \textbf{5}, 1259 (1957). 


\bibitem{DMI2} T. Moriya, Anisotropic superexchange interaction and weak ferromagnetism, Phys. Rev. \textbf{120}, 91 (1960).






%---------multiferroics spiral------------------



\bibitem{yamasaki_PRL_2007}
Y. Yamasaki, H. Sagayama, T. Goto, M. Matsuura, K. Hirota, T. Arima, and Y. Tokura, Electric control of spin helicity in a magnetic ferroelectric, Phys. Rev. Lett. \textbf{98}, 147204 (2007).

\bibitem{Ramesh_PRL_2022} E. Parsonnet, L. Caretta, V. Nagarajan, H. Zhang, H. Taghinejad, P. Behera, X. Huang, P. Kavle, A. Fernandez, D. Nikonov, H. Li, I. Young, J. Analytis, and R. Ramesh, Nonvolatile electric field control of thermal magnons in the absence of an applied magnetic field, Phys. Rev. Lett. \textbf{129}, 087601 (2022).

\bibitem{Ramesh_NM_2024} X. Huang, X. Chen, Y. Li, J. Mangeri, H. Zhang, M. Ramesh, H. Taghinejad, P. Meisenheimer, L. Caretta, S. Susarla, R. Jain, C. Klewe, T. Wang, R. Chen, C. H. Hsu, I. Harris, S. Husain, H. Pan, J. Yin, P. Shafer, Z. Qiu, D. R. Rodrigues, O. Heinonen, D. Vasudevan, J. Iniguez, D. G. Schlom, S. Salahuddin, L. W. Martin, J. G. Analytis, D. C. Ralph, R. Cheng, Z. Yao, and R. Ramesh, Manipulating chiral spin transport with ferroelectric polarization, Nat. Mater. \textbf{23}, 898 (2024).


\bibitem{Ramesh_NC_2024} P. Meisenheimer, G. Moore, S. Zhou, H. Zhang, X. Huang, S. Husain, X. Chen, L. W. Martin, K. A. Persson, S. Griffin, L. Caretta, P. Stevenson, and R. Ramesh, Switching the spin cycloid in BiFeO$_3$ with an electric field, Nat. Commun. \textbf{15}, 2903 (2024).


\bibitem{Ramesh_AM_2024} P. Meisenheimer, M. Ramesh, S. Husain, I. Harris, H. W. Park, S. Zhou, H. Taghinejad, H. Zhang, L. W. Martin, J. Analytis, P. Stevenson, J. \'{I}\~niguez-Gonz\'alez, S. K. Kim, D. G. Schlom, L. Caretta, Z. Yao, and R. Ramesh, Designed spin-texture-lattice to control anisotropic magnon transport in antiferromagnets, Adv. Mater. \textbf{36}, 2404639 (2024).


\bibitem{Ramesh_npj_2025} S. Husain and R. Ramesh, Magnon transport in multiferroic antiferromagnet, npj Spintronics \textbf{3}, 36 (2025).


\bibitem{Ramesh_PRL_2025} I. A. Harris, S. Husain, P. Meisenheimer, M. Ramesh, H. W. Park, L. Caretta, D. G. Schlom, Z. Yao, L. W. Martin, J. {\'I}{\~n}iguez-Gonz{\'a}lez, S. W. Kim, and R. Ramesh, Symmetry-Based Phenomenological Model for Magnon Transport in a Multiferroic, Phys. Rev. Lett. \textbf{134}, 016703 (2025).


\bibitem{NC-BFO-torque} Y. Liang, X. Yan, X. Jiang, D. Jiang, H. Bai, Z.-A. Wang, H. Chen, X. Qiu, D. Pharis, X. Huang, R. Jain, D. Shao, W. Jiang, D. Yi, D. C. Ralph, and T. Nan, Unconventional magnon-mediated spin torque enabled by ferroelectric domain engineering in multiferroic BiFeO$_3$, Nat. Commun. \textbf{17}, 5140 (2026).


\bibitem{nonlocal} L. J. Cornelissen, J. Liu, R. A. Duine, J. Ben Youssef, and B. J. van Wees, Long-distance transport of magnon spin information in a magnetic insulator at room temperature, Nat. Phys. \textbf{11}, 1022 (2015).



\bibitem{Gross2017} I. Gross, W. Akhtar, V. Garcia, L. J. Mart{\'i}nez, S. Chouaieb, K. Garcia, C. Carr{\'e}tero, A. Barth{\'e}l{\'e}my, P. Appel, P. Maletinsky, J.-V. Kim, J. Y. Chauleau, N. Jaouen, M. Viret, M. Bibes, S. Fusil, and V. Jacques, Real-space imaging of non-collinear antiferromagnetic order with a single-spin magnetometer, Nature \textbf{549}, 252 (2017).



\bibitem{Gitgeatpong2017} G. Gitgeatpong, Y. Zhao, P. Piyawongwatthana, Y. Qiu, L. W. Harriger, N. P. Butch, T. J. Sato, and K. Matan, Nonreciprocal Magnons and Symmetry-Breaking in the Noncentrosymmetric Antiferromagnet, Phys. Rev. Lett. \textbf{119}, 047201 (2017).



\bibitem{Chen2015Dissipationless} W. Chen and M. Sigrist, Dissipationless Multiferroic Magnonics, Phys. Rev. Lett. \textbf{114}, 157203 (2015).


\bibitem{supplement} See Supplemental Material at [...] for the detailed derivation of magnon dispersion and magnon spin matrix for the magnetic spirals, discussions of the parity of linear polarization, and a detailed setup of the quantum kinetic equation for magnon diffusive transport. 



%-------AFM and ATM------

\bibitem{offdiagonal_magnon_spin} N. Okuma, Magnon spin-momentum locking: Various spin vortices and Dirac magnons in noncollinear antiferromagnets, Phys. Rev. Lett. \textbf{119}, 107205 (2017).

\bibitem{spin_matrix} X. Ye and T. Yu, Magnon correlation enables spin injection, dephasing, and transport in canted antiferromagnets, Phys. Rev. B \textbf{112}, 224417 (2025).



\bibitem{AFM1} S. Seki, T. Ideue, M. Kubota, Y. Kozuka, R. Takagi, M. Nakamura, Y. Kaneko, M. Kawasaki, and Y. Tokura, Thermal generation of spin current in an antiferromagnet, Phys. Rev. Lett. \textbf{115}, 266601 (2015).



\bibitem{AFM2} W. Lin, K. Chen, S. Zhang, and C. L. Chien, Enhancement of thermally injected spin current through an antiferromagnetic insulator, Phys. Rev. Lett. \textbf{116}, 186601 (2016).



\bibitem{AFM3} J. Li, Z. Shi, V. H. Ortiz, M. Aldosary, C. Chen, V. Aji, P. Wei, and J. Shi, Spin Seebeck effect from antiferromagnetic magnons and critical spin fluctuations in epitaxial FeF$_2$ films, Phys. Rev. Lett. \textbf{122}, 217204 (2019).




\bibitem{AFM4} J. Li, C. B. Wilson, R. Cheng, M. Lohmann, M. Kavand, W. Yuan, M. Aldosary, N. Agladze, P. Wei, M. S. Sherwin, and J. Shi, Spin current from sub-terahertz-generated antiferromagnetic magnons, Nature (London) \textbf{578}, 70 (2020).



\bibitem{AFM5} P. Vaidya, S. A. Morley, J. van Tol, Y. Liu, R. Cheng, A. Brataas, D. Lederman, and E. del Barco, Subterahertz spin pumping from an insulating antiferromagnet, Science \textbf{368}, 160 (2020).



\bibitem{ATM1} Q. Cui, B. Zeng, P. Cui, T. Yu, and H. Yang, Efficient spin Seebeck and spin Nernst effects of magnons in altermagnets, Phys. Rev. B \textbf{108}, L180401 (2023).



\bibitem{ATM2} Y. Lei, R.-C. Xiao, W. Lin, and T. Yu, Electrical injection and control of vector spin flow in altermagnets, Phys. Rev. Research \textbf{8}, 043122 (2026).



\bibitem{AFM_plane1} A. J. Sievers and M. Tinkham, Far infrared antiferromagnetic resonance in MnO and NiO, Phys. Rev. \textbf{129}, 1566 (1963).



\bibitem{AFM_plane2} S. J. Williamson and S. Foner, Antiferromagnetic resonance in systems with Dzyaloshinsky-Moriya coupling; Orientation dependence in $\alpha$-Fe$_2$O$_3$, Phys. Rev. \textbf{136}, A1102 (1964).


\bibitem{AFM_plane3} X. Ye, Q. Cui, W. Lin, and T. Yu, Spin quenching and transport by hidden Dzyaloshinskii-Moriya interactions, Phys. Rev. B \textbf{111}, 064401 (2025).



\bibitem{Bauer2012} G. E. W. Bauer, E. Saitoh, and B. J. van Wees, Spin caloritronics, Nat. Mater. \textbf{11}, 391 (2012).


\bibitem{s-d_interaction_1} S. S.-L. Zhang and S. Zhang, Spin convertance at magnetic interfaces, Phys. Rev. B \textbf{86}, 214424 (2012).


\bibitem{s-d_interaction_2} N. Rohling, E. L. Fj{\ae}rbu, and A. Brataas, Superconductivity induced by interfacial coupling to magnons, Phys. Rev. B \textbf{97}, 115401 (2018).


\bibitem{s-d_interaction_3} E. Thingstad, E. Erlandsen, and A. Sudb{\o}, Eliashberg study of superconductivity induced by interfacial coupling to antiferromagnets, Phys. Rev. B \textbf{104}, 014508 (2021).


\bibitem{s-d_interaction_4} P. Tang and G. E. W. Bauer, Electrical injection and transport of coherent magnon superposition states in noncollinear antiferromagnets, Phys. Rev. B \textbf{112}, L180406 (2025).


\bibitem{s-d_interaction_5} K. Sourounis and A. Manchon, Efficient generation of spin currents in altermagnets via magnon drag, Phys. Rev. B \textbf{111}, 134448 (2025).


\bibitem{Yu2013} T. Yu and M. W. Wu, Spin relaxation in an ultracold spin-orbit-coupled $^{40}$K gas, Phys. Rev. A \textbf{88}, 043634 (2013).


\bibitem{relaxation-time1} S. Das, A. Ross, X. X. Ma, S. Becker, C. Schmitt, F. van Duijn, E. F. Galindez-Ruales, F. Fuhrmann, M.-A. Syskaki, U. Ebels, V. Baltz, A.-L. Barra, H. Y. Chen, G. Jakob, S. X. Cao, J. Sinova, O. Gomonay, R. Lebrun, and M. Kl\"aui, Anisotropic long-range spin transport in canted antiferromagnetic orthoferrite YFeO$_3$, Nat. Commun. \textbf{13}, 6140 (2022).


\bibitem{relaxation-time2} H. Wang, R. Yuan, Y. Zhou, Y. Zhang, J. Chen, S. Liu, H. Jia, D. Yu, J.-P. Ansermet, C. Song, and H. Yu, Long-distance coherent propagation of high-velocity antiferromagnetic spin waves, Phys. Rev. Lett. \textbf{130}, 096701 (2023).

\end{thebibliography}
\end{document}